# Multiplexed spin-wave-photon entanglement source using temporal-multimode memories and feed-forward-controlled readout

*Yafei Wen, Pai Zhou, Zhongxiao Xu, Liang Yuan, Haoyi Zhang, Shengzhi Wang, Long Tian, Shujing Li, and Hai Wang\**

*The State Key Laboratory of Quantum Optics and Quantum Optics Devices, Institute of Opto-Electronics, Collaborative Innovation Center of Extreme Optics, Shanxi University, Taiyuan 030006, People's Republic of China*

A source that can generate atom-photon quantum correlations or entanglement based on a quantum memory is a basic building block of quantum repeaters (QRs). To achieve high entanglement generation rates in ensemble-based QRs, spatial-, temporal- and spectral-multimode memories are needed. Previous temporal-multimode memories are based on rephasing mechanisms in inhomogeneously broadened media. Here, by applying a train of write pulses in time, with each pulse coming from a different direction, to a homogeneously broadened atomic ensemble to induce Duan-Lukin-Cirac-Zoller-like Raman processes, we prepare up to 19 pairs of modes, namely, one spin-wave mode and one photonic time bin. Spin-wave-photon (i.e., atom-photon) entanglement is probabilistically produced in these mode pairs. Based on the stored spin-wave modes together with feed-forward-controlled readout, we build a temporally multiplexed source and then demonstrate an 18.8-fold increase in the probability of the generation of spin-wave-photon entanglement compared to the sources that use individual modes. The measured Bell parameter for the multiplexed source is $2.30 \pm 0.02$, and the memory lifetime is 30 $\mu s$.

## I. Introduction

The distribution of entanglement over long distances is critical for quantum communications [2-3] and quantum networks [4], however, it is difficult because of unavoidable transmission losses. Quantum repeater (QR) [1] holds promise for overcoming this problem. In the QR protocol, a long distance is divided into short elementary links. Entanglement is generated independently in each link and then extended to the full distance by entanglement swapping. For a practical realization of a QR, Duan *et al*. presented a proposal called the Duan-Lukin-Cirac-Zoller (DLCZ) protocol [2]. This protocol has been highly influential. Building on this protocol, several improved QR protocols have been proposed [5-10]. In these DLCZ-like protocols, the entanglement generation in each link relies on atomic-ensemble-based sources that may probabilistically generate quantum correlation or entanglement between a photon and a spin-wave excitation (a quantum memory) [2-8]. To suppress multiple excitations, the probabilities of preparing the spin-wave-photon entanglement (SWPE) or quantum correlation are kept low. In these cases, a large number of attempts are needed to generate entanglement in every link [3]. An advantage of the protocols that use entanglement sources [7], compared to those that use quantum correlation sources [2, 5], is that long-distance phase stability is no longer required [3]. Over the past ~ 20 years, sources have been demonstrated by using spontaneous Raman scattering in cold atomic ensembles [11-28] or storing one member of an entangled photon pair in a solid-state [29-31] or gas-state atomic ensemble [32]. Unfortunately, the quantum memories in these sources are single-mode. When using these sources in a QR, only one entanglement creation attempt per link per communication time interval $L_0/c$ can be achieved [3], where $L_0$ is the separation distance between the two ends (nodes) of the link, and $c$ is the light speed in optical fibers. The communication time interval $L_0/c$ is very long, for example, for a typical link of $L_0 = 60 km$, $L_0/c \approx 300 \mu s$. Such a long interval $L_0/c$ leads to very slow rates for the entanglement generation in each QR link [3]. To enhance the repeater rate, QR schemes based on temporal- [33-34], spectral- [35-36], and spatial-multimode [37-40] memories have been proposed. Recently, a review paper emphasized that multimode operations are necessary for practically useful repeaters [41]. In the initial temporal-multiplexing QR scheme [33], a photon pair source and a memory capable of storing independent $N$ temporal modes are used to build the QR node, and then the entanglement creation rate per link per $L_0/c$ is increased by a factor of $N$. Compared with the other schemes, temporal-multiplexing schemes are attractive since they repeatedly use the same physical process and then reduce the resources. A controlled rephasing mechanism in an inhomogeneously broadened medium, such as an atomic frequency comb (AFC) [42] or gradient echo [43], was proposed to realize a temporal-multimode memory. Based on this mechanism, the storage of multiple light pulses has been experimentally demonstrated in solid-state ensembles [44-47] and atomic vapors [43, 48-49]. Recently, quantum correlations between a photon and a spin-wave excitation in more than ten modes were demonstrated in crystals using the AFC spin-wave scheme [50-52]. Given that the most advanced experiments on QRs have been performed in atomic gases, C. Simon *et al.* proposed a temporally multiplexed version of the DLCZ protocol using controlled rephasing of inhomogeneous spin broadening [34]. Following this proposal, a spin-wave-photon quantum correlation in two modes was demonstrated with a cold atomic ensemble [53]. Thus far, the generation of temporal-multimode spin-wave-photon correlations in a homogeneously broadened medium has remained unexplored.

Here, by applying a train of write pulses in time, with each pulse coming from a different direction, to a homogeneously broadened atomic ensemble to drive spontaneous Raman transitions, we prepare $m$ mode pairs (MPs) of a spin-wave mode and a photonic time bin (temporal mode). The $m$ spin-wave modes are spatially stored in the ensemble in a distinguishable manner, and $m$ time bins propagate in a given spatial mode. The entanglement between one Stokes photon and one spin-wave excitation is probabilistically created in these MPs. The detection of a Stokes photon in the $i$-th time mode will herald the storage of an excitation in the *i*-th spin-wave mode. With feed-forward-controlled readout, only the heralded excitation is retrieved. The memory modes are finally emptied by a clean pulse. For a write-clean cycle, a multiplexed source using *m*=19 spin-wave modes increases the SWPE generation probability by a factor of ~18.8 compared to nonmultiplexed sources that use the individual modes. The measured Bell parameter for the multiplexed source is $2.30\pm0.02$, and the memory lifetime is 30 $\mu s$.

## II. Multiplexed scheme and experimental setup.

The experimental setup for the multiplexed source is shown in Fig. 1. The atomic ensemble is a cloud of cold $^{87}$Rb atoms [Fig. 1(a)] that is loaded by a magneto-optical trap and has an optical density of approximately 9. The atomic ground levels $|a\rangle$ and $|b\rangle$ together with the excited level $|e_1\rangle$ ($|e_2\rangle$) form a $\Lambda$-type configuration [Fig. 1(b)]. After the atoms are released from the magneto-optical trap, we prepare the atoms into Zeeman levels $|a,m_{F_a}\rangle$ ($m_{F_a}=0,\pm1$) and then start the SWPE generation. In the beginning of a trial [Fig. 1(c)], a train of laser write pulses is applied to the atoms to create spin-wave memories and Stokes-photon emissions. The train contains $m$ write pulses $w(t_i)$ ($i=1$ to $m$) and lasts $\Delta T=7\mu s$, with $t_i$ being the times at which the $w(t_i)$ pulses are applied ($t_1=0$). The *m* write pulses go through the center of the ensemble along *m* different directions, corresponding to *m* wavevectors $k_{wi}$ ($i=1$ to $m$). All the write pulses are $\sigma^+$–polarized and blue-detuned from the $|a\rangle \to |e_2\rangle$ transition by 10 MHz, which spontaneously induces Raman transitions $|a, m_{F_a}\rangle \to |b, m_{F_b} = m_{F_a}\rangle$ and $|a, m_{F_a}\rangle \to |b, m_{F_b} = m_{F_a}+2\rangle$ via $|e_1, m_{F_{e1}} = m_{F_a}+1\rangle$. The transition $|a, m_{F_a}\rangle \to |b, m_{F_b} = m_{F_a}\rangle$ ($|a, m_{F_a}\rangle \to |b, m_{F_b} = m_{F_a}+2\rangle$) induced by the write pulse $w(t_i)$ may emit a $\sigma^+$ ($\sigma^-$)-polarized Stokes photon $|R\rangle_{S(t_i)}$ ($|L\rangle_{S(t_i)}$) into the time bin $S(t_i)$ and simultaneously create a spin-wave excitation $|1\rangle^+_{M(t_i)}$ ($|1\rangle^-_{M(t_i)}$) in mode $M(t_i)$. The atom-photon joint state created by $w(t_i)$ may be written as $\rho^{t_i}_{ap} = |0\rangle^{(t_i)(t_i)}\langle 0| + \chi_i |\Phi\rangle^{(t_i)(t_i)}_{ap\ ap}\langle\Phi|$, where $|0\rangle^{(t_i)} = |0\rangle_{S(t_i)}|0\rangle_{M(t_i)}$ is the vacuum part, $|0\rangle_{S(t_i)}$ ($|0\rangle_{M(t_i)}$) denotes the state without a photon (excitation) in $S(t_i)$ ($M(t_i)$), $\chi_i$ ($\ll 1$) is the excitation probability, $|\Phi\rangle^{t_i}_{ap} = (\cos\vartheta |1^+_{M(t_i)}\rangle|R_{S(t_i)}\rangle + \sin\vartheta |1^-_{M(t_i)}\rangle|L_{S(t_i)}\rangle)$ is the SWPE state created in the $i$th MP, and $\cos\vartheta$ is the relevant Clebsch-Gordan coefficient. In the present experiment, the excitation probabilities for various $M(t_i)$ modes are basically symmetric, i.e., $\chi_1 \approx ... \chi_i ... \approx \chi_m \approx \chi$. The Stokes photons in all the time bins $S(t_i)$ are collected by a single-mode fiber $SMF_S$; thus, the photons have the wavevector $k_S$. The wavevector of the spin-wave excitation in the $M(t_i)$ mode is given by $k_{M_i} = k_{W_i} - k_S$. For an ensemble atomic number of $N \gg 1$, the $m$ spin-wave modes created at times $t_1, t_2, ... t_m$ satisfy the orthogonal relations $\langle 1^-_{M(t_j)}|1^-_{M(t_i)}\rangle \approx \delta_{\mathbf{k}_{M_i}\mathbf{k}_{M_j}}$ and $\langle 1^+_{M(t_j)}|1^+_{M(t_i)}\rangle \approx \delta_{\mathbf{k}_{M_i}\mathbf{k}_{M_j}}$ [54]; thus, the modes are distinguishably stored in the ensemble. After the $SMF_S$, we transform the $\sigma^+(\sigma^-)$-polarization of the Stokes photon into a horizontal (vertical)-polarization by using a $\lambda/4$ plate. Then, the Stokes photons are guided to a polarizing beam splitter ($PBS_S$), which transmits the horizontal (*H*) polarization to a detector $D_1$ and reflects the vertical (*V*) polarization to a detector $D_2$. As shown in Fig. 1(c), the gates of the detectors $D_{1,2}$ are opened during the time bins $S(t_1)...S(t_i)..S(t_m)$. If a Stokes photon is detected by the detector $D_1$ or $D_2$ in one of these time bins, e.g., in time bin $S(t_l)$, the storage of one spin-wave excitation in the mode $M(t_l)$ is heralded, which corresponds to the $|\Phi\rangle^{ap}_{(t_l)}$ state being created. This photon detection event at $D_1$ or $D_2$ is processed by a field-programmable gate array (FPGA) and is then included in the Stokes detection probability $P^{(m)}_{S1}$ or $P^{(m)}_{S2}$. It should be noted that during the application of the write pulse train, if two Stokes photons are detected in the $S(t_l)$ and $S(t_k)$ time bins ( $l,k \in m$ and $l<k$ ), only the detection event in $S(t_l)$ is registered by the FPGA and then counted into $P^{(m)}_{S1}$ or $P^{(m)}_{S2}$. After a storage time of $\tau$, the FPGA outputs a feed-forward signal. Controlled by the signal, a reading laser $R_l$ with direction $k_{R_l}=-k_{w_l}$ and frequency tuned to the $|b\rangle \to |e_1\rangle$ transition is switched on, which converts the excitation in the $M(t_l)$ mode into an anti-Stokes photon. Due to the atomic collective interference, the anti-Stokes photons retrieved from various modes $M(t_i)$ ( $i=1$ to $m$ ) are efficiently coupled to the spatial mode $A_s$, which is along the direction $k_{A_s}=k_{W_l}-k_S-k_{R_l}=-k_S$ . We use a single-mode fiber $SMF_A$ to collect the anti-Stokes photon in $A_s$ (see Fig. 1). After the $SMF_A$, the $\sigma^+$ ($\sigma^-$) -polarized anti-Stokes photon is transformed into the *H* ($V$)-polarization by a $\lambda/4$ wave-plate. The two-photon entanglement state transformed from $|\Phi\rangle_{ap}^{t_i}$ can be written as $|\Phi\rangle_{pp}^{t_i} = (\cos\vartheta |H\rangle_{A_s}|H\rangle_{S_i} + \sin\vartheta |V\rangle_{A_s}|V\rangle_{S_i})$, where $|H\rangle_{A_s}$ ($|V\rangle_{A_s}$) and $|H\rangle_{S_i}$ ($|V\rangle_{S_i}$) denote an $H$ ($V$)-polarized photon in the mode $A_s$ and a photon in the mode $S(t_i)$, respectively. The anti-Stokes photons in the mode $A_s$ then impinge on a polarizing beam splitter ($PBS_A$), which transmits $H$-polarized photons (reflects $V$-polarized photons) to a detector $T_1$ ($T_2$). The gates of $T_{1,2}$ are opened during bin $A_s$ [Fig. 1(c)]. If an anti-Stokes photon is detected by $T_1$ or $T_2$, a coincident count between $D$ ($D_1$ or $D_1$) and $T$ ($T_1$ or $T_2$) is successful and then is included in Stokes-anti-Stokes coincidence count rate (CCR) denoted by $C_{D_1 T_1}^{(m)}(X,Y)$, $C_{D_1 T_2}^{(m)}(X,Y)$, $C_{D_2 T_1}^{(m)}(X,Y)$ and $C_{D_2 T_2}^{(m)}(X,Y)$, where $X$ and $Y$ denote the measured polarization settings of the Stokes and anti-Stokes fields, respectively. After the retrieval (transformation), a cleaning laser pulse controlled by the FPGA is applied to empty the memories, and then the atoms are pumped back to the initial state $|a\rangle$. If a Stokes photon is not detected during the time bins $S(t_1)$, ..., $S(t_m)$, the retrieval is not performed, and only the cleaning pulse will be applied to pump the atoms back to the initial state $|a\rangle$. After the cleaning pulse, the trial (write-clean cycle) ends, and the next trial starts.

The multiplexed source may generate SWPE in $m$ modes and then increase the generation probability per write-clean cycle by a factor of $m$ compared to the nonmultiplexed source, which generates the SWPE in a single mode. Such an increase can be translated into an increase in the entanglement generation rate of a QR elementary link [3, 33]. We explain this result based on the ideas in Refs. [3, 33] below. To build an elementary link, one may use a pair of nonmultiplexed or multiplexed SWPE sources, with each placed at a node of the link. A trial of the entanglement generation starts by applying two synchronized write pulses (for the nonmultiplexed case) or write pulse trains (for the multiplexed case) to the source's ensembles. The Stokes photons from the two sources are sent to a middle station between the nodes for a Bell-state measurement (BSM). The spin-wave excitations must be stored until the BSM's result is sent back to the memories in the two nodes. If the BSM is not successful, each memory has to be emptied by a cleaning pulse, and the next trial starts. The time interval per write-clean cycle (trial) includes the time intervals of the write and cleaning processes ($\delta t_w$ and $\delta t_c$) and the communication time $L_0/c$ [33, 34]. For the presented multiplexed and nonmultiplexed sources, $\delta t_w$ and $\delta t_c$ are all in the range of $0.1-10\mu s$, which are much smaller than the communication time $L_0/c \approx 300\mu s$ for a typical link ($L_0 = 60km$), and therefore can be neglected. Thus, the time intervals per trial for the multiplexed and nonmultiplexed links are all the communication time $L_0/c$. For a communication time $L_0/c$, the multiplexed link allows $m$ entanglement attempts, while the nonmultiplexed link allows one attempt. Thus, the multiplexed link will give rise to an *m*-fold increase in the entanglement generation rate compared to the nonmultiplexed link.

## III. Experimental results and analysis

The generation probability of the SWPE can be evaluated by the Stokes detection probability measured in the *X* =*H-V* polarization basis per write-clean run, which is written as $P_S^{(1)} = P_{S1}^{(1)} + P_{S2}^{(1)} = \chi\eta_D$ ( $P_S^{(m)} = P_{S1}^{(m)} + P_{S2}^{(m)}$ ) for the nonmultiplexed (multiplexed) source; $P_{S1}^{(1)}$ and $P_{S2}^{(1)}$ ( $P_{S1}^{(m)}$ and $P_{S2}^{(m)}$ ) are the probabilities of detecting a photon by D1 and D2 in the individual time bin, for example, the 1$^{st}$ time bin (any one of the *m* time bins) per write-clean run, respectively. For the present experiment, where the excitation probabilities for the individual sources are all $\chi$, the Stokes detection probabilities for the individual time bins may all be written as $P_S^{(1)} = \chi\eta_D$. For $\chi\eta_D < 1$, the generation probability of the multiplexed source is $P_S^{(m)} = 1-(1-P_S^{(1)})^m \approx mP_S^{(1)}$ [33, 38]. Therefore, by observing $P_S^{(m)}$ as a function of *m*, we can demonstrate that the capacity of the multiplexed source may increase the SWPE generation probability. The blue square dots in Fig. 2 are the measured values of $P_S^{(m)}$ as a function of *m* in the *H-V* basis. The measured results show that $P_S^{(m)}$ increases linearly with the increase in *m*. From the measured data for $P_S^{(m=19)}$ and $P_S^{(1)}$ in Fig. 2, we obtain $P_S^{(m=19)} / P_S^{(1)} \approx 18.8$, which is in agreement with the expected value of *m*=19.

As mentioned above, for a multiplexed source, if a Stokes photon is detected in a time bin, for example, in the *i*-th bin, the storage of an excitation in the *i*-th spin wave is heralded. After the storage time, the excitation can be individually converted into an anti-Stokes photon via feed-forward-controlled retrieval. Such feed-forward-controlled retrieval is required for entanglement swapping between adjacent links in a multiplexed QR [33, 35], enabling the increase in the entanglement generation rate of a link to be translated into an increase in the repeater rate by the same factor. The Stokes and anti-Stokes photons are in an entangled state with a controllable delay, which may be written as $|\Phi\rangle_{pp}^{m} = (\cos\vartheta|H\rangle_A|H\rangle_S + \sin\vartheta|V\rangle_A|V\rangle_S)$, where $|H\rangle_S$ ($|V\rangle_S$) denotes an $H$ ($V$)-polarized Stokes photon in one of the $S(t_i)$ ($i=1$ to $m$) modes, and $|H\rangle_A$ ($|V\rangle_A$) denotes an $H$ ($V$)-polarized anti-Stokes photon in the mode $A_S$.

The quality of the entangled state $|\Phi\rangle_{pp}^{m}$ can be characterized by the Bell parameter $S^{(m)} = |E^{(m)}(\theta_S, \theta_A) - E^{(m)}(\theta_S, \theta_A') + E^{(m)}(\theta_S', \theta_A) + E^{(m)}(\theta_S', \theta_A')| < 2$, with the correlation function $E^{(m)}(\theta_S, \theta_A)$ given by $\frac{C^{(m)}_{D_1 T_1}(\theta_S,\theta_A) + C^{(m)}_{D_2 T_2}(\theta_S,\theta_A) - C^{(m)}_{D_1 T_2}(\theta_S,\theta_A) - C^{(m)}_{D_2 T_1}(\theta_S,\theta_A)}{C^{(m)}_{D_1 T_1}(\theta_S,\theta_A) + C^{(m)}_{D_2 T_2}(\theta_S,\theta_A) + C^{(m)}_{D_1 T_2}(\theta_S,\theta_A) + C^{(m)}_{D_2 T_1}(\theta_S,\theta_A)}$, where $\theta_S$ and $\theta_S'$ ($\theta_A$ and $\theta_A'$) are the polarization bases of the Stokes (anti-Stokes) field, which are set by rotating the plate $WP_S$ ($WP_A$) before the $PBS_S$ ($PBS_A$). In the measurement, the canonical settings are chosen to be $\theta_S = 0°$, $\theta_S' = 45°$, $\theta_A = 22.5°$ and $\theta_A' = 67.5°$. To investigate the multimode storage ability of the ensemble, we measure the decay of the Bell parameter $S^{(m=19)}$ with the storage time $\tau$ for $\chi \approx 1\%$. The blue square dots in Fig. 3 depict the measured $S^{(m=19)}$ data. For $\tau = 700ns$,

$S^{(m=19)} = 2.30 \pm 0.02$, while at $\tau = 30 \mu s$, $S^{(m=19)} = 2.03 \pm 0.02$, which violate the Clauser-Horne-Shimony-Holt (CHSH) inequality by 15 and 1.5 standard deviations, respectively.

The entanglement quality of the 19-mode multiplexed source can also be characterized by the fidelity of the state $|\Phi\rangle_{pp}^{(m=19)}$, which is defined by $F^{(m=19)} = \left(Tr\sqrt{\sqrt{\rho_r^{(m=19)}} \rho_d \sqrt{\rho_r^{(m=19)}}}\right)^2$, where $\rho_r^{(m=19)}$ ($\rho_d$) is the reconstructed (ideal) density matrix of the entangled state. By measuring the Stokes-anti-Stokes CCRs $C_{D_1 T_1}^{(m)}(X,Y)$, $C_{D_1 T_2}^{(m)}(X,Y)$, $C_{D_2 T_1}^{(m)}(X,Y)$ and $C_{D_2 T_2}^{(m)}(X,Y)$ for the $X$ and $Y=H(V)$, $D(A)$, and $R(L)$ polarization bases, which are set by using wave plates $WP_S$ and $WP_A$ (see Fig. 1), respectively, we reconstruct $\rho_r^{(m=19)}$ and plot the result in Fig. 4, which yields $F^{(m=19)} = 0.859 \pm 0.003$.

The standard DLCZ-like source uses only a single-mode memory to generate the SWPE, where its Bell parameter is up to ~2.6 [11] for an excitation probability of $\chi \approx 1\%$, which is obviously larger than the presented result of ~2.30. To understand this result, we measure the Bell parameters of the SWPE sources with varying $m$. The measured Bell parameter $S^{(m)}$ as a function of $m$ is plotted in Fig. 5 (blue circle dots), which shows that $S^{(m)}$ is equal to $2.65 \pm 0.03$ for $m=1$ and decreases as $m$ increases. We attribute this decrease to background noise, which is simply explained in the following. When we apply a train containing $m$ write pulses to the ensemble to prepare the spin-wave modes, a large number of unwanted spin waves that are associated with undetected

Stokes photons are also created. When a read pulse is applied to retrieve the heralded spin wave, the unwanted spin waves may also be converted into anti-Stokes photons. Due to phase mismatch of the conversion processes, the photons are nondirectional emissions [see Appendix A], which leads to background noise and then degrades the SWPE quality. This issue is similar to that addressed in Ref. [32], in which the noise results from the out-of-phase spin waves and may be overcome by using a moderate-finesse cavity resonant with the Stokes photons but off-resonant with the anti-Stokes photons.

The probability of generating the polarization-entangled photon pair per write-clean run can be evaluated by the Stokes-anti-Stokes coincidence probability $P_{S,AS}^{(m)}$ measured in the *H-V* basis. The multiplexed source uses feed-forward-controlled readout, which may convert the heralded spin-wave excitation into an anti-Stokes photon. Thus, the coincidence probability may be expressed as $P_{S,AS}^{(m)} = \sum_{i=1}^{m}\left((P_{S1}^{(ith)} + P_{S2}^{(ith)})\gamma_{ii}\eta_{AS}\right)$, where $\eta_{AS}$ is the efficiency of the anti-Stokes channel, and $\gamma_{ii}$ is the retrieval efficiency for the *i*-th mode. In the present experiment, the retrieval efficiencies for various modes are basically symmetric ($\gamma_{11} \approx ...\gamma_{ii}... \approx \gamma_{mm} \approx \gamma$), such that we have $P_{S,AS}^{(m)} \approx m\chi\eta_A\gamma\eta_{AS} = mP_{S,AS}^{(1)}$, where $P_{S,AS}^{(1)} = \chi\eta_A\gamma\eta_{AS}$ is the probability of generating a polarization-entangled photon pair from the nonmultiplexed sources. The above result indicates that the increase in the SWPE probability translates into an increase in the probability of the entangled two-photon pair. The red square dots in Fig. 5 show $P_{S,AS}^{(m)}$ as a function of *m*, which are obtained by measuring the Stokes-anti-Stokes CCRs $C_{D_1 T_1}^{(m)}(X,Y)$ and $C_{D_2 T_2}^{(m)}(X,Y)$ for the *X*, *Y=H-V* polarization basis and using the relation $P_{S,AS}^{(m)} = \left( C_{D_1 T_1}^{(m)}(H\text{-}V, H\text{-}V) + C_{D_2 T_2}^{(m)}(H\text{-}V, H\text{-}V) \right) / r_m$, where $r_m$ is the repetition rate, which is kept constant at $4.6 \times 10^4 / s$ for the different values of *m*. From the measured data in Fig. 5, we obtain $P_{S,AS}^{(m=19)} / P_{S,AS}^{(1)} \approx 18.3$, which is consistent with the expected value of *m*=19.

**IV. Comparisons of our presented experiment with two previous works**

Before we conclude, we would like to compare our presented experiment with two previous works [13, 14]. The two experiments also utilized a DLCZ-like quantum memory to demonstrate the probability increases in the generation of a spin-wave excitation. Such increases rely on a fast feedback protocol and a single-mode quantum memory, which are suitable for building deterministic and storable single-photon sources. The generation probability of a single excitation conditioned on detecting a Stokes photon per write pulse is $\eta_S \chi$ for the single-mode memory. To increase the generation probability, the previous experiments applied a series of trials with a period of $\delta t_{w/c}$ to the atoms. Each trial contained a write pulse and a cleaning pulse, forming a write-cleaning cycle. Once a Stokes photon is detected by a detector in a trial, for example, in the *j*-th trial, a feedback signal will be sent out, and further trials will be stopped. At a predetermined time $T$, a read pulse is applied to the atoms to convert the excitation into a single photon. The time interval $T$ is limited by the memory lifetime $\tau_0$ (i.e., $T \leq \tau_0$), which gives a maximum number of trials $N = T/\delta t_{w/c} \approx \tau_0/\delta t_{w/c}$. A feedback protocol with $N$ trials applied to the atoms may generate a single excitation according to the new probability $\sum_{i=0}^{N-1} \eta\chi(1-\eta\chi)^i \approx N\eta\chi$ ($\eta\chi \ll 1$). For deterministically generating a single excitation, one may apply $N \sim 1/\eta\chi$ trials to the atoms, which requires that the memory lifetime is longer than the time of $T \approx \delta t_{w/c}/\eta\chi$. The single excitation can be converted into a single photon at the predetermined time $T$. Such single-photon generation will be synchronized with that of the other memories and then may be used to generate local multiphoton states [55, 56]. The generation rate is limited by the time interval $T = N\delta t_{w/c}$ and then by $\delta t_{w/c}$ for a given value of $N$. The period $\delta t_{w/c}$ mainly includes the write and cleaning pulse durations and the transmission time of the Stokes photon to the detector and the feedback delay. To build deterministic and storable single-photon sources, local generations of single excitations are involved, which require a small spatial separation between the atoms and the detector and then leads to a fast feedback loop. The value of the period $\delta t_{w/c}$ is $0.3\mu s$ in [14] and $1\mu s$ in [13], respectively, and can be further decreased by decreasing the write (cleaning) pulse duration.

The presented temporally multiplexed memory may also be used for single excitation generation. A multiplexed memory capable of storing $m$ modes may enhance the generation probability from $\eta\chi$ to $m\eta\chi$ per write-clean cycle, where the write-clean cycle is formed by a train of write pulses followed by a clean pulse. Such enhanced generation requires a time of $\delta t_{w/c}^{(m)}$, which is equal to the period of a write-clean cycle for the multiplexed case. Similar to the single-mode case, the required time $\delta t_{w/c}^{(m)}$ mainly includes the durations of the write pulse train and cleaning pulses. This time is limited by the lifetime $\tau_0$ of the multimode memory ($\delta t_{w/c}^{(m)} \leq \tau_0$). Due to the use of the write pulse train, $\delta t_{w/c}^{(m)}$ is longer than $\delta t_{w/c}$. To evaluate $\delta t_{w/c}^{(m)}$ from $\delta t_{w/c}$, we need their relationship. For simplicity, we assume that a train containing $m$ write pulses may be formed by removing the cleaning pulses from $m$ write-clean cycles for the single-mode protocol. In this case, we have a rough relationship of $\delta t_{w/c}^{(m)} \approx m\delta t_{w/c}$. The enhanced generation with a probability of $m\eta\chi$ can also be obtained by using the single-mode feedback protocol with $N = m$ trials, which require the same time of $m\delta t_{w/c}$ as $\delta t_{w/c}^{(m)}$. Thus, when the lifetime of the single-mode memory is the same as that of the multimode memory, the achieved maximal generation probabilities using both memories are symmetric. However, the experiment for the single-mode protocol is easier than that for the multiplexed memory protocol. Thus, the single-mode feedback protocol is more appropriate for enhancing the generation probability of the single excitation.

Temporally multiplexed SWPE sources apply well to the QR, which involves entanglement distribution over a long distance. The QR protocol divides a long distance into short elementary links and requires that the entanglement is generated independently in each link. A trial of entanglement generation in the link requires the application of two synchronized write-clean cycles to the two ensembles. As mentioned above, the required times per write-clean cycle for the multiplexed and the nonmultiplexed links are all equal to the communication time $L_0/c$. For the write process, the link using the multiplexed SWEP sources allows $m$ entanglement attempts, while the link using the nonmultiplexed SWEP sources allows one attempt. Thus, the multiplexed link will give rise to an $m$-fold increase in the entanglement generation probability compared to that of the nonmultiplexed link, which can be described by $P_L^{(m)} \approx m P_L^{(1)}$, where $P_L^{(1)}$ and $P_L^{(m)}$ denote the entanglement generation probabilities per nonmultiplexed ($m$-mode multiplexed) link. In the QR protocol, it is required that entanglement is successfully generated in parallel in each link. Once entanglement is generated in a link, it is stored in quantum memories while waiting for entanglement generation in the adjacent links. Since the probability $P_L^{(1)}$ ($P_L^{(m)}$) will be less than unity for the presented cases, one has to repeat the trial many times to successfully generate entanglement in the nonmultiplexed (multiplexed) link. The required maximum number of trials and the average time for successful entanglement generation in the single-mode ($m$-mode) link are $\frac{1}{P_L^{(1)}}$ ($\frac{1}{mP_L^{(1)}}$) and $\frac{L_0}{c}\frac{1}{P_L^{(1)}}$ ($\frac{L_0}{c}\frac{1}{mP_L^{(1)}}$), respectively. One can see that the use of $m$-mode memory will decrease the required waiting time (lifetime) for the entanglement generation in a link by a factor of $m$ compared to that of a single-mode memory. Thus, the repeater rate will be increased by using multimode memories.

## V. Conclusion

We demonstrate a new scheme capable of creating temporal-multimode SWPE in a homogeneously broadened atomic ensemble. By multiplexing 19 memory modes, we experimentally demonstrate the expected increases in the entanglement generation probabilities, which represents a key step towards the realization of temporally multiplexed QRs. It should be noted that the presented temporally multiplexed memory is different from the previous spatially multiplexed memory [36]. The parallel developments of both memories are critical for achieving a dramatic increase in the memory capacity. For example, if $m$ temporal modes and $n$ spatial modes are combined into a system, one can achieve $m \times n$ memory modes, promising a tremendous increase in the SWPE generation probability and then effectively improving the repeater rate. The background noise generated in the multimode preparation degrades the quality of the SWPE. However, this is not a fundamental issue. A moderate-finesse ($F$) cavity can be used to enhance the Stokes emissions into the cavity mode by a factor of $2F/\pi$ [25] and then reduce the background noise by the same factor [32]. In our present experiment, the memory lifetime is only $\sim 30\,\mu s$, which is limited by the dephasing effects resulting from the atomic motions and inhomogeneous broadening of the spin transition [3]. By loading the atoms into an optical lattice [20-23] and selecting two magnetic-field-insensitive spin waves to store memory qubits [21, 57], the lifetime of the quantum memory will be improved by more than 2 orders of magnitude. We believe that the presented scheme paves a way for improving the entanglement distribution rate of long-distance quantum communication.

**Appendix A. Nondirectional emissions induced by unwanted spin waves in the multiplexed case**

When a train of write pulses is applied to the ensemble to create the desired spin waves $M(t_i)$, which are paired with the time bins $S(t_i)$ ($i=1,2,...m$) propagating along the $SMF_S$ direction (z-axis in Fig. 6), a large number of unwanted spin waves, which are associated with the Stokes photons propagating along the directions different from the z-axis, will be probabilistically created. In the feed-forward-controlled retrieval, the unwanted spin waves may be converted into nondirectional emissions.

We now consider an example to explain this result. In an experimental write process, we assume that a Stokes photon along the z-axis is detected in a time bin, for example, in the *l*-th time bin ($S(t_l)$), and an undetected Stokes photon propagating along the direction at an angle $\theta$ (see Fig. 6) relative to the *z*-axis is created in another time bin, for example, in the *k*-th bin ($l \neq k$). The Stokes photon $S(t_l)$ is paired with the desired spin wave $M(t_l)$. The Stokes photon in the *k*-th bin and the unwanted spin wave are denoted as $S(t_k,\theta)$ and $M(t_k,\theta)$, respectively. The spin-wave $M(t_k,\theta)$ has a wavevector of $k_M(t_k,\theta) = k_{w_k} - k_s(t_k,\theta)$, where $k_{w_k}$ is the wavevector of the write pulse $w(t_k)$, and $k_s(t_k,\theta)$ is the wavevector of the Stokes photon $S(t_k,\theta)$. In the present experiment, the frequencies of the write, read, Stokes and anti-Stokes light fields are basically identical, which are denoted as $\omega$. When the read pulse $R_l$ is applied to retrieve the spin wave $M(t_l)$, the excitation in the spin-wave mode $M(t_k,\theta)$ will also be converted into an anti-Stokes photon $A_S(\theta_A)$. The angle between the propagating directions of the write pulse $w(t_k)$ (read pulse $R_l$) and the z-axis is assumed to be $\theta_{w_k}$ ($\theta_{R_l}$) (see Fig. 6), and the wavevectors of the pulses $w(t_k)$, $R_l$ and the Stokes photon $S(t_k,\theta)$ may be expressed as $k_w(t_k) = |k|\cos\theta_{w_k} + i|k|\sin\theta_{w_k}$, $k_{R_l} = -|k|\cos\theta_{R_l} - i|k|\sin\theta_{R_l}$, and $k_s(t_k,\theta) = |k|\cos\theta + i|k|\sin\theta$, respectively, where, $|k| = \omega/c$. The phase matching condition (PMC) requires that the wavevector of the anti-Stokes photon $A_S(\theta_A)$ should be $k_{A_s}(\theta_{A_s}) = k_{w_g} + k_{R_l} - k_s(t_g,\theta)$, i.e., $k_{A_s}(\theta_A) = |k|\left((\cos\theta_{w_k} - \cos\theta_{R_l} - \cos\theta) + i(\sin\theta_{w_k} - \sin\theta_{R_l} - \sin\theta)\right)$. For $k \neq l$, we calculate $(\cos\theta_{w_k} - \cos\theta_{R_l} - \cos\theta)^2 + (\sin\theta_{w_k} - \sin\theta_{R_l} - \sin\theta)^2 \neq 1$, i.e., $|k_{A_s}(\theta_A)| \neq |k|$, which indicates that the emission of the anti-Stokes photon $A_s(\theta_A)$ does not satisfy the PMC and thus is nondirectional.

## Acknowledgments

We acknowledge funding support from Key Project of the Ministry of Science and Technology of China (Grant No. 2016YFA0301402), the National Natural Science Foundation of China (Grants: No. 11475109, No. 11274211, No. 11604191, No. 11804207 and No. 61805133), Shanxi "1331 Project" Key Subjects Construction and Program for Sanjin Scholars of Shanxi Province.

*Corresponding author: wanghai@sxu.edu.cn

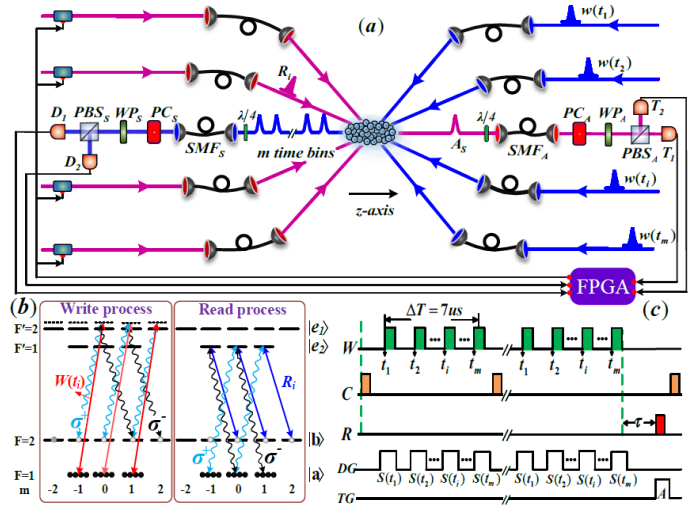

FIG. 1. (a) Experimental setup for $m$-mode multiplexed sources. The pulse trains, each containing $m$ write pulses, are applied to the ensemble along different directions to generate the SWPE, with $m$ being up to 19 (for simplicity, we only plot 4-direction write pulses). $PC_{S,A}$: phase compensators; $WP_{S,A}$: half-wave or quarter-wave plates. In the measurements of the fidelities in Fig. 4, $WP_{S,A}$ are half-wave (quarter-wave) plates

when analyzing the photon polarization in the *DA (RL)* polarization setting and are removed for the *HV* polarization setting. In the measurements of the Bell parameters in Figs. 3 and 5, $WP_{S,A}$ are half-wave plates and used to set the polarization angles. (b) Relevant atomic levels. (c) Time sequence of the experimental trials. *W, C, R*: write, cleaning, and read pulses. *DG (TG)*: Timeline of the *D (T)* detector gate; the sizes of the time bins $S(t_1), S(t_2)... S(t_m)$ and $A_S$ are all ~70 ns.

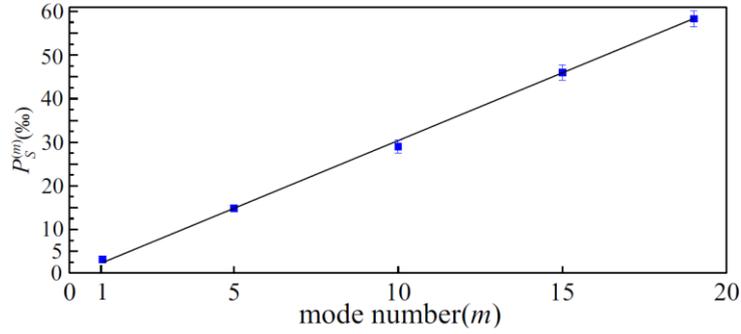

FIG. 2. The measured Stokes detection probability $P_S^{(m)}$ as a function of the mode number *m*. The solid line is a linear fit to the measured data.

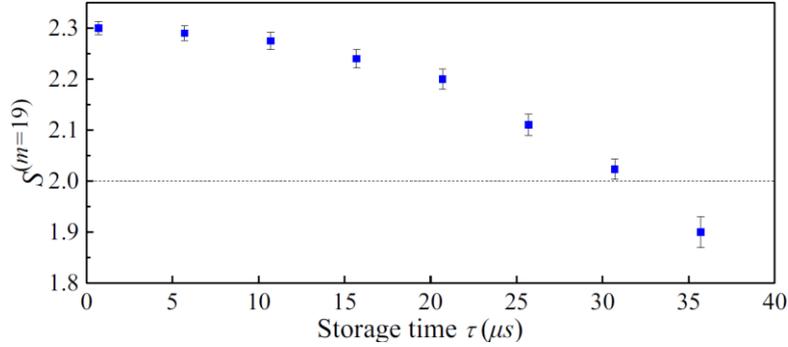

FIG. 3. Measurements of the Bell parameter $S^{(m=19)}$ as a function of $\tau$ for $\chi = 1\%$. The error bars represent 1 standard deviation.

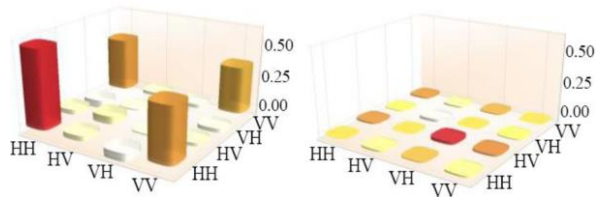

FIG. 4. Real and imaginary parts of the reconstructed density matrices of the two-photon entangled state $|\Phi\rangle_{pp}^{(m=19)}$ for $\tau = 700\,ns$ and $\chi = 1\%$.

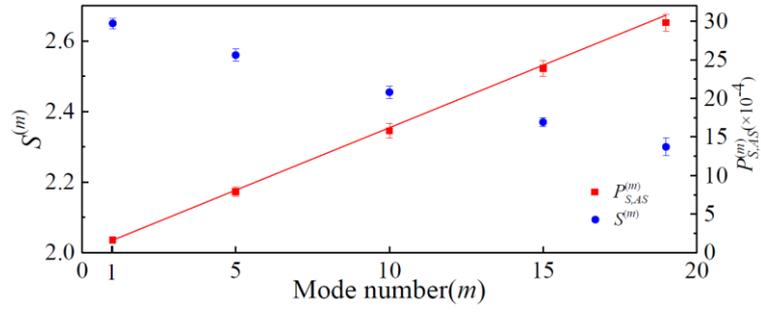

FIG. 5. Measured Bell parameter $S^{(m)}$ and the Stokes-anti-Stokes coincidence probability $P_{S,AS}^{(m)}$ as a function of $m$ for $\tau = 700\,ns$ and $\chi = 1\%$. The red solid line is a linear fit to the measured $P_{S,AS}^{(m)}$ data.

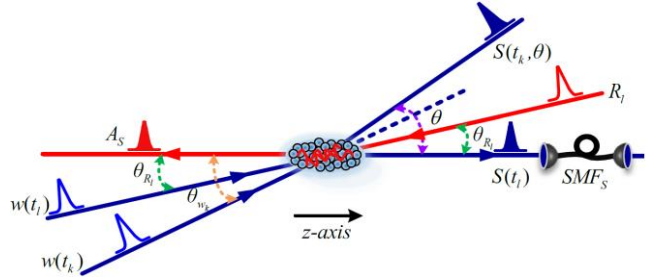

FIG. 6. The relative propagation directions of the write and read pulses and the desired and unwanted Stokes time bins.